%% file: main_active_T1.tex
\documentclass[a4paper,twoside,11pt,aps,superscriptaddress,amsmath,amsfonts,amssymb,pre,floatfix,longbibliography,nofootinbib]{revtex4-1}

\pdfoutput=1
 
\usepackage{amsmath}
\usepackage{amssymb}
\usepackage{bbold} 

\usepackage{color}
\usepackage[dvipsnames]{xcolor}
\definecolor{nblue}{RGB}{28,130,185}
\definecolor{cgreen}{RGB}{76,153,0}
\definecolor{bostonuniversityred}{rgb}{0.8, 0.0, 0.0}
\definecolor{myorange}{RGB}{245,156,74}

\newcommand{\new}[1]{\textcolor{black}{#1}}

\usepackage{hyperref}
\hypersetup{
  colorlinks=true,
  citecolor=magenta,
  urlcolor=-myorange
}

\usepackage[utf8]{inputenc}
\usepackage{subfigure}
\usepackage{xifthen}
\usepackage{graphicx}
\usepackage{hyphenat}
\usepackage{hhline}
\usepackage{upgreek}
\usepackage{bm}
\usepackage{xcolor}

\usepackage{multirow}

\usepackage{subfiles} 

\newcommand{\DDt}[1]{%
\frac{{\rm D}#1}{{\rm D}t}%
}
\newcommand{\DDtinline}[1]{%
{\rm D}#1/{\rm D}t%
}

\newcommand{\ddt}[1]{%
\frac{{\rm d}#1}{{\rm d}t}%
}

\begin{document}

\title{Active T1 transitions in cellular networks}

\author{Charlie~\surname{Duclut}}
\thanks{These authors contributed equally.}
\affiliation{Max Planck Institute for the Physics of Complex Systems, N\"othnitzer Str. 8, 01187 Dresden, Germany}
\author{Joris~\surname{Paijmans}}
\thanks{These authors contributed equally.}
\affiliation{Max Planck Institute for the Physics of Complex Systems, N\"othnitzer Str. 8, 01187 Dresden, Germany}
\author{Mandar~M.~\surname{Inamdar}}
\affiliation{Department of Civil Engineering, Indian Institute of Technology Bombay, Powai, Mumbai 400076, India}
\author{Carl~D.~\surname{Modes}}
\affiliation{Max Planck Institute for Molecular Cell Biology and Genetics (MPI-CBG), Dresden 01307, Germany}
\affiliation{Center for Systems Biology Dresden, Pfotenhauerstrasse 108, 01307 Dresden, Germany}
\affiliation{Cluster of Excellence, Physics of Life, TU Dresden, Dresden 01307, Germany}
\author{Frank~\surname{J\"ulicher}}
\affiliation{Max Planck Institute for the Physics of Complex Systems, N\"othnitzer Str. 8, 01187 Dresden, Germany}
\affiliation{Center for Systems Biology Dresden, Pfotenhauerstrasse 108, 01307 Dresden, Germany}
\affiliation{Cluster of Excellence, Physics of Life, TU Dresden, Dresden 01307, Germany}

\begin{abstract}
In amorphous solids as in tissues, neighbor exchanges can relax local stresses and allow the material to flow. In this paper, we use an anisotropic vertex model to  study T1 rearrangements in polygonal cellular networks. We consider two different  physical realizations of the active anisotropic stresses: (i) anisotropic bond tension and (ii) anisotropic cell stress.
Interestingly, the two types of active stress lead 
\new{to patterns of relative orientation of T1 transitions and cell elongation that are different. Our work suggests that these two realizations of anisotropic active stresses can be observed \textit{in vivo}.}
We describe and explain these results through the lens of a continuum description of the tissue as an anisotropic active material. We furthermore discuss the energetics of the dynamic tissue and express the energy balance in terms of internal elastic energy, mechanical work, chemical work and heat. This allows us to define active T1 transitions that can perform mechanical work while consuming chemical energy. 
\end{abstract}

\maketitle


\section*{Introduction}

During morphogenesis, complex structures emerge starting from a single fertilized egg as the results of the collective organization of a large number of cells.
Understanding principles that govern self-organization of cells into complex structures and organs is one of the major challenges of biology and biophysics.
The collective behavior of cells relies on chemical signals~\cite{hogan1999morphogenesis,ben-jacob2000cooperative,mahdisoltani2021nonequilibrium,tzur2009cell}, but also depends on cellular force generation and active mechanical processes as well as tissue mechanical properties~\cite{heisenberg2013forces,wyatt2016question}.
Morphogenesis, i.e., the generation of shape, is therefore a result of self-organized processes that couple chemical signalling with mechanical activity~\cite{guirao2015unified,etournay2015interplay,iyer2019epithelial,dye2021selforganized}.

Change in tissue shape involves anisotropic active processes and cell rearrangements.
The physics of tissue dynamics is based on a description of tissues as active viscoelastic and viscoplastic materials~\cite{popovic2021inferring,mongera2018fluidtosolid,tlili2020migrating,kim2021embryonic}.
Depending on timescales, tissues can behave like solids, able to withstand external shear stresses, or like fluids, and can rearrange their cells and exhibit cell flows~\cite{mongera2018fluidtosolid,tlili2020migrating,kim2021embryonic}. 
Such rearrangements permit the maintenance of mechanical integrity of a tissue while changing local connectivity and the overall shape. 
In tissues, rearrangements can result from cell divisions or extrusions, where new cells are added or removed from the tissue, which has been shown to permit tissue fluidization~\cite{ranft2010fluidization}. In addition, cells can also rearrange and change neighbors in so-called T1 transitions.

T1 transitions have been studied first in passive materials such as foams, where they occur in response to external shear forces that can drive material flow~\cite{cohen-addad2013flow,biance2011how}.
Tissues, however, are active materials, that can deform spontaneously, driven by internally generated stresses, and can therefore also perform work on their environment.
Such active deformations are for instance observed during convergence–extension, a widespread morphogenetic process driven by oriented T1 transitions that leads to anisotropic tissue deformation~\cite{keller2000mechanisms,tada2012convergent}.
In contrast to passive foams, where T1 transitions dissipate energy and relax elastic stresses resulting from external forcing, T1 transitions in tissues can be active and perform work, and therefore can build up stresses rather than relaxing them.
The orientation of T1 transitions can be guided by tissue polarity cues, that are linked to chemical signals such as the planar polarity pathways~\cite{wang2007tissue,bosveld2012mechanical}. Thereby, tissues can extend along axes that are defined by chemical patterns.
Such processes can be observed in developmental model systems. For example, during the germ-band extension of \textit{Drosophila} embryo, experiments show that the tissue deforms anisotropically \new{as a consequence of oriented T1 transitions, driven by active processes in the acto-myosin cytoskeleton. Two scenarios have been proposed: anisotropic accumulation of myosin II at cell-cell junctions~\cite{bertet2004myosindependent,rauzi2008nature,tetley2016unipolar,wang2020anisotropy}, and anisotropic active stresses mediated by medial myosin pulses~\cite{collinet2015local}.}
Similarly, data from the pupal wing blade of \textit{Drosophila} reveal multiple roles of T1 transitions over time: they drive anisotropic cell and tissue elongation at early stages, while later they are responsible for a relaxation of cell shape elongation~\cite{etournay2015interplay}.
\new{Finally, cells can actively propel themselves on a substrate. Such motion can also cause T1 transitions and tissue shape changes~\cite{lin2018dynamic,barton2017active,sussman2017cellgpu}, and can trigger a solid-to-fluid transition in cell tissues model~\cite{bi2016motilitydriven}. Here, we focus of anisotropic stresses generated in a tissue in the absence of active self-propulsion.}

We use a two-dimensional vertex model to study how anisotropic tissue stresses can drive oriented cell rearrangements and anisotropic tissue shape changes.
In particular, we discuss active T1 transitions that can perform mechanical work, in contrast to passive T1 transitions that relax elastic stresses.
Vertex models provide simple models of tissue physics that can capture cell shape, packing geometry and cell rearrangements~\cite{etournay2015interplay,alt2017vertex,fletcher2014vertex,tetley2019tissue,comelles2021epithelial,bi2015densityindependent,bi2016motilitydriven,yamamoto2020nonmonotonic}.
\new{The role of anisotropic internal stresses has been studied within the vertex model framework, for instance using anisotropy in cell bond tension~\cite{wang2020anisotropy,rauzi2008nature}, or by introducing cell bond tension that depends on the identity of adjacent cells~\cite{tetley2016unipolar}.}
Following Ref.~\cite{duclut2021nonlinear}, we consider a cell network where a preferred axis is set by a nematic field that represents tissue polarity. We discuss two different physical realizations of anisotropic active stress: (i) anisotropic bond tension, where the contractility of bonds is increased along a preferred axis, and (ii) anisotropic cell stress, where the bulk of the cells exhibits an anisotropic stress that is contractile along a preferred axis.
Surprisingly, we find that these two realizations, although involving an anisotropic active stress along the same direction, lead to cell rearrangements and cell elongation patterns which are very different. 
\new{Our analysis suggests that the early stages of \textit{Drosophila} pupal wing morphogenesis could be an example where anisotropic bond tension dominates, while both realizations of anisotropic active stress could contribute during germ-band extension in the \textit{Drosophila} embryo. In the latter case, anisotropic cell stress could be a consequence of medial myosin II pulses that favors the opening of newly-formed cell bonds~\cite{collinet2015local}.}
We complement our analysis and understanding by using a coarse-grained continuum description of the cell network. This description uses concepts from active matter theory~\cite{marchetti2013hydrodynamics,prost2015active,julicher2018hydrodynamic} and has proven valuable to characterize the large-scale properties of tissues~\cite{popovic2017active,alert2019active,perez-gonzalez2019active,duclut2019fluid,duclut2021hydraulic}.
By considering the energetics of the cell network, we show that active stresses can induce T1 transitions that perform mechanical work. We call these \textit{active T1 transitions}.

The paper is organized as follows. In Sec.~\ref{sec_vertex_model}, we present the vertex model and its modification to account for anisotropic bond tension and cell stress. We then introduce a linear anisotropic continuum model to capture tissue dynamics. In Sec.~\ref{sec_results_active_T1}, we quantify the outcome of anisotropic vertex model simulations and highlight the differences between the two implementation of anisotropy. Fits of the continuum model to the simulation results provide us with a better understanding of the mechanisms at play. We finally discuss the energetics of the tissues, allowing us to provide a definition of active T1 transitions. 

\section{Mechanics of anisotropic cell networks}

\label{sec_vertex_model}

\sloppy

\begin{figure}[t]
 \centering
 \includegraphics[width=1.0\textwidth]{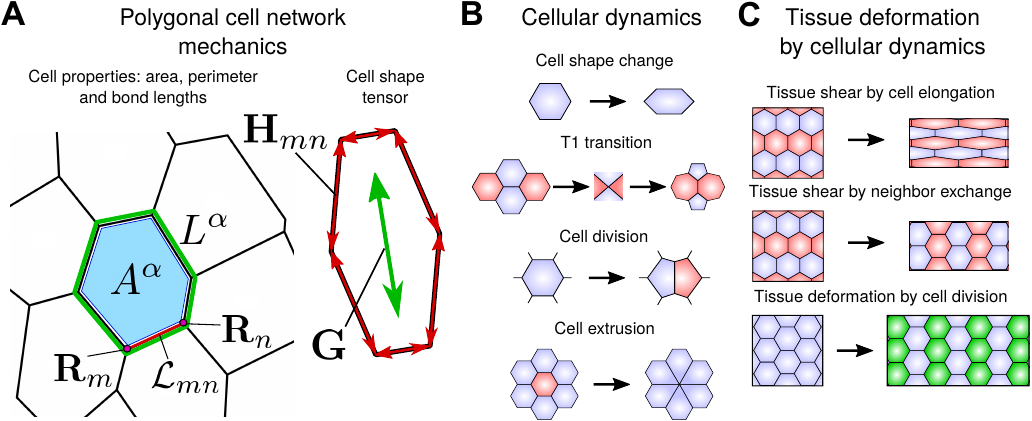}
 \caption{
Mechanics and dynamics of cellular networks. 
\textbf{(A)} Definition of the cell state variables. 
Left shows the cell area $A^\alpha$ (blue patch), cell perimeter $L^\alpha$ (green line) and bond length $\mathcal{L}_{mn}$ (red line) between the vertices with positions $\mathbf{R}_m$ and $\mathbf{R}_{n}$.
Right shows the cell elongation tensor $\mathbf{G}$ which is constructed from the bond nematic tensors $\mathbf{H}_{mn}$ as defined in Eq.~\eqref{cell_shape_tensor}. 
\textbf{(B)} Cell dynamic processes can lead to tissue deformation as an effect of cell shape changes, T1 transitions, cell divisions or cell extrusions.
\textbf{(C)} Large-scale tissue deformation can be driven by collective cell dynamics: 
cell shape changes (top), anisotropic T1 transitions (middle) and anisotropic cell divisions (bottom). The tissue may also deform as a result of changes in the mean cell shape of the cellular network.
 }
 \label{fig_polygonal_cell_network}
\end{figure}

The apical junctions of an epithelial tissue can be described by a packing of convex polygons and its mechanics can be described by a vertex model, where cells are represented as polygons that are outlined by straight edges connecting vertices~\cite{farhadifar2007influence}. 
We consider a polygonal cell network consisting of $N_{\rm c}$ cells. Each cell $\alpha$ is characterized in terms of its area $A^\alpha$, its perimeter $L^\alpha$ and the lengths $\mathcal{L}_{mn}$ of the bonds that form the outline of the cell, where $m$ and $n$ label the vertices that they connect (see Fig.~\ref{fig_polygonal_cell_network} for illustration). 

We employ a quasistatic representation of epithelia where the cell network is at any instant in a mechanical equilibrium, while the parameters describing cell properties can slowly change with time.
At each vertex $m$, the total force $\mathbf{F}_m = -\partial E_0/\partial \mathbf{R}_m$ vanishes, where $\mathbf{R}_m$ is the position of the vertex, and $E_0$ is the vertex model work function and reads~\cite{honda1984computer, farhadifar2007influence}:
\begin{equation}
  E_0 = \sum_\alpha \frac{1}{2} K^\alpha\left( A^\alpha - A^\alpha_0 \right)^2 + \sum_{\langle m,n \rangle} \Lambda_{mn} \mathcal{L}_{mn}+ 
  \sum_\alpha \frac{1}{2} \Gamma^\alpha (L^\alpha)^2.
 \label{eq_tissue_work}
\end{equation}
Note that for clarity, upper-case letters are used here and in the following for quantities related to the vertex model, while lower-case letters will be used for the continuum model.
The first term describes an area elasticity contribution, with $A^\alpha_0$ the preferred cell area and $K^\alpha$ the area stiffness.
The second term describes a contribution due to the tension of network bonds with length $\mathcal{L}_{mn}$ and line tension $\Lambda_{mn}$. 
The third term describes an elasticity of the cell perimeter with stiffness~$\Gamma^\alpha$.

Nonequilibrium dynamics of the vertex model are captured by a time-dependent line tension~$\Lambda_{mn}(t)$. 
The line tension dynamics of individual bonds in the network follows an Ornstein--Uhlenbeck process:
\begin{equation}
 \ddt{\Lambda_{mn}} = -\frac{1}{\tau_\Lambda}(\Lambda_{mn}(t) - \bar\Lambda_{mn}) + \Delta\Lambda \sqrt{2/\tau_\Lambda} \, \Xi_{mn}(t) \, ,
 \label{eq_line_tension_dyn_isotropic}
\end{equation}
where $\Xi_{mn}(t)$ is a Gaussian white noise with zero mean $\langle \Xi_{mn}(t) \rangle = 0$, and correlations \mbox{$\langle \Xi_{mn}(t)\Xi_{op}(t') \rangle = \delta_{\langle mn \rangle,  \langle op \rangle} \delta(t-t')$} 
where $\delta_{\langle mn \rangle,  \langle op \rangle}=1$ if bonds $\langle mn \rangle $ and $\langle op \rangle$ are the same and 0 otherwise~\cite{farhadifar2007influence, aigouy2010cell}. The line tension of every bond relaxes toward its mean value $\bar\Lambda_{mn}$ with a characteristic time $\tau_\Lambda$, which sets the timescale of the dynamics and is of the order of the acto-myosin cortex turn-over time.

As discussed for instance in Refs.~\cite{duclut2021nonlinear,yamamoto2020nonmonotonic}, the magnitude of bond tension fluctuations $\Delta\Lambda$ has a crucial role in the rheological properties of cell networks. A low value of this fluctuation magnitude leads to a glassy dynamics and non-linearities dominate. In the following, we are interested in a regime where $\Delta\Lambda$ is sufficiently large, such that the vertex model has linear viscoelastic properties.

A polygonal network described by the work function~\eqref{eq_tissue_work} has isotropic properties. In the following, we consider how this description can be extended to describe anisotropic cell networks.

    \subsection{Anisotropic cellular networks}

Motivated by planar cell polarity in tissues~\cite{wang2007tissue,bosveld2012mechanical}, we consider that the anisotropy of the network can be described by a unit nematic field $\bm{\mathcal{P}}$ assigned to each polygon and which gives locally a preferred axis.
In two dimensions, the nematic field $\bm{\mathcal{P}}$ can be parameterized by a single angle~$\Psi$ which defines the direction of the anisotropy axis (see App.~\ref{sec_nematic_tensor_decomposition} for details). 
For simplicity, we consider in the following that $\Psi$ is constant and thus provides a global preferred axis in the tissue.

\medskip 

\noindent\textbf{Anisotropic bond tension.}
To include the effect of such a nematic field in the dynamics of the vertex model, we first consider an anisotropic bond tension, such that its mean magnitude~$\bar\Lambda_{mn}$ depends on the orientation of the bond with respect to the nematic field~$\mathcal{P}$. We choose 
\begin{align}
    \bar\Lambda_{mn} = \bar\Lambda_{mn}^0 \left( 1 + \beta \bm{\mathcal{P}}: \hat{\mathbf{H}}_{mn} \right) \, ,
    \label{eq_anisotropic_bond_tension}
\end{align}
where $\bm{A}:\bm{B}= {\rm Tr}(\bm{A}\cdot\bm{B})$ denotes the full tensor contraction, and where $\beta \geq 0$ is dimensionless and sets the magnitude of the anisotropy. We have introduced the unit bond nematic $\hat{\mathbf{H}}_{mn}=\mathbf{H}_{mn}/\mathcal{L}_{mn}^2$ between vertices $m$ and $n$. Here, $\mathbf{H}_{mn}$ is the nematic tensor constructed from the vector $\bm{\mathcal{L}}_{mn}$ pointing from vertex $m$ to vertex $n$ as \mbox{$\mathbf{H}_{mn}=\bm{\mathcal{L}}_{mn} \otimes \bm{\mathcal{L}}_{mn} - \mathcal{L}_{mn}^2 \mathbb{1}/2$} with $\mathbb{1}$ the unit tensor in two dimensions (see Fig.~\ref{fig_polygonal_cell_network}).
With this definition, a bond making an angle $\Phi$ with the local nematic field has a mean bond tension that reads:
\begin{align}
    \bar\Lambda_{mn} = \bar\Lambda_{mn}^0 \left( 1 + \beta \cos(2\Phi) \right) \, .
\end{align}
Consistently with our convention for the anisotropic cell stress, the definition given by Eq.~\eqref{eq_anisotropic_bond_tension} with $\beta>0$ implies that cells have a higher bond tension along the axis set by $\bm{\mathcal{P}}$. As a consequence, they are more likely to undergo a T1 transition along this axis. 

\medskip 

\noindent\textbf{Anisotropic cell stress.} 
An alternative description of anisotropic tissues can be obtained by considering an anisotropic cell stress $\bm\Sigma^{\mathrm{a}} = \Sigma^{\mathrm{a}} \bm{\mathcal{P}}$, where $\Sigma^{\mathrm{a}}$ is the magnitude of the active stress. The work performed by this anisotropic stress is added to the vertex model work function as\footnote{Note that we are using a different sign convention compared to Ref.~\cite{duclut2021nonlinear}.}:
\begin{align}
  E = E_0 + \sum_\alpha \frac{1}{2}A^\alpha \bm{\Sigma}^{\mathrm{a}} : \bm{G}^\alpha \, ,
 \label{eq_tissue_work_anisotropic}
\end{align}
where $\bm{G}^\alpha$ is the cell shape tensor of each cell $\alpha$
\begin{align}
\bm G^\alpha = \frac{1}{A^\alpha}\sum_{\langle m,n \rangle} \mathbf{H}_{mn} \, .
 \label{cell_shape_tensor}
\end{align}
The cell shape tensor $\bm G^\alpha$ quantifies the deviation of the cell shape from isotropic shapes, for which $\bm G^\alpha$ vanishes.
See Fig.~\ref{fig_polygonal_cell_network} for an illustration.
With the definition of Eq.~\eqref{eq_tissue_work_anisotropic} and $\Sigma^{\rm a}>0$, the anisotropic cell stress implies a stronger contractility of the cells along the direction set by $\bm{\mathcal{P}}$, and cells therefore tend to elongate in the direction orthogonal to $\bm{\mathcal{P}}$.

Note that for simplicity, we use in the following the same constant values of the parameters $K^\alpha$, $A^\alpha_0$, $\Gamma^\alpha$ for all cells and the same value $\bar\Lambda_{mn}^0=\bar\Lambda^0$ for all bonds.
\new{Note that the presence of bond tension fluctuations prevents crystalization of the cellular pattern that could be otherwise observed in a monodisperse system~\cite{durand2019thermally,li2018role}.}
In App.~\ref{sec_vertex_model_implementation}, we give details on the numerical implementation of the vertex model. Values of the (dimensionless) parameters used in the simulations are given in Table~\ref{tab_vertex_model_parameters}.

   \subsection{Dynamics of a polygonal cell network and shear decomposition}
    \label{sec_dynamics}

The deformation of a cellular network is quantified by its shear rate, which can be decomposed into cellular contributions. For flat polygonal networks, such a decomposition can be done exactly~\cite{etournay2015interplay, merkel2017triangles}.
Following Ref.~\cite{merkel2017triangles}, the large-scale shear-rate tensor $\widetilde V_{ij}$ of the cellular network can be decomposed as: 
\begin{align}
    \widetilde V_{ij} = \frac{{\rm D} Q_{ij}}{{\rm D}t} + R_{ij} \, .
    \label{eq_shear_decomp}
\end{align}
Here and in the following, $i$ and $j$ corresponds to 2d Cartesian indices, $Q_{ij}$ is the mean cell elongation tensor and ${\rm D}/{\rm D}t$ is the corotational time derivative of a tensor (defined in Eq.~\eqref{eq_SI_hydr_model_corotational_derivative_simple_shear} of App.~\ref{sec_definitions}). The tensor $R_{ij}$ accounts for shear rate due to topological rearrangements and is a sum of four contributions:
\begin{align}
    R_{ij} = T_{ij} +  C_{ij} +  E_{ij} +  D_{ij} \, ,
    \label{eq_shear_decomp_2}
\end{align}
where the tensors $T_{ij}$, $C_{ij}$ and $E_{ij}$ account for shear rate due to T1 transitions, cell divisions and cell extrusions, respectively.
The tensor $D_{ij}$ is a shear rate 
associated with heterogeneities and fluctuations. If such fluctuations are correlated, they contribute to shear even if they vanish on average.  In particular, the tensor $D_{ij}$ includes shear stemming from correlations between triangle rotations and triangle elongation as well as correlations between triangle area changes and triangle elongation~\cite{etournay2015interplay,merkel2017triangles}. 
Note finally that all the tensors introduced in Eqs.~\eqref{eq_shear_decomp} and~\eqref{eq_shear_decomp_2} are two-dimensional \textit{nematic} tensors. It means that they are symmetric traceless tensors 
\new{that define an orientation and a magnitude.}
They are fully characterized by two independent quantities: a norm and an angle with respect to the $x$-axis (see also App.~\ref{sec_nematic_tensor_decomposition}). 

Note that the trace of the velocity gradient tensor $V_{kk}$ (summation over repeated indices is implied), which corresponds to isotropic tissue growth, can also be decomposed into cellular contributions~\cite{merkel2017triangles,popovic2017active}. Here, we only focus on the anisotropic contributions.
Finally, the tissue stress tensor $\Sigma_{ij}$ in the simulations is symmetric and can be decomposed into an isotropic pressure and a symmetric traceless part, the shear stress $\widetilde \Sigma_{ij}$.

    \subsection{Hydrodynamic model for cellular networks under anisotropic active stress}

The viscoelastic behavior of stochastic cellular networks can be captured by a continuum model of tissues~\cite{marmottant2009role, etournay2015interplay, popovic2017active,duclut2021nonlinear}. Such a coarse-grained description does not hold at a single-cell level but requires an averaging over many cells, as provided by the shear decomposition~\eqref{eq_shear_decomp} of a triangulated network discussed above.   

In the continuum description, we therefore introduce the anisotropic part of the deformation rate tensor $\tilde v_{ij}$, which can be decomposed into cellular contributions due to changes in the mean cell elongation tensor $q_{ij}$ and shear $r_{ij}$ caused by topological rearrangements. Note that we use lower-case letters for the continuum model description. We therefore have:
\begin{subequations}%
\begin{equation}%
 \tilde v_{ij} = \DDt{q_{ij}} + r_{ij} \, ,
 \label{eq_cont_shear_decomp}
\end{equation}
where $\DDtinline{}$ denotes the corotational derivative defined in Eq.~\eqref{eq_SI_hydr_model_corotational_derivative_simple_shear}. 

We also include in our continuum description the fact that the axis of topological rearrangements is biased both by the axis of cell elongation and the axis of active anisotropic processes. This fact is captured by introducing linear relationships between the shear contribution from topological rearrangements~$r_{ij}$, the cell elongation~$q_{ij}$, and the anisotropic axis $p_{ij}$. It reads~\cite{popovic2017active,duclut2021nonlinear}:
\begin{align}
 r_{ij} = \frac{1}{\tau} \, q_{ij} + \lambda p_{ij} \, ,
 \label{eq_r_constitutive}
\end{align}
where $\tau$ is the characteristic timescale of topological rearrangements and $\lambda$ is the rate of anisotropic cell rearrangements.

We also introduce the tissue stress $\sigma_{ij}$, which we decompose into an isotropic part and an anisotropic symmetric traceless part, the tissue shear stress $\tilde \sigma_{ij}$.
We consider that the cellular network is an active elastic material, such that to linear order, the shear stress can be written as:
\begin{align}
    \tilde \sigma_{ij} = \mu \, q_{ij} + \zeta p_{ij} \, ,
    \label{eq_sigma_constitutive}
\end{align}%
\label{eq_continuum_model}%
\end{subequations}%
where $\mu$ is the shear modulus of the tissue and $\zeta$ is the anisotropic active stress magnitude. 

\section{Cell elongation and T1 transitions driven by active processes}

\label{sec_results_active_T1}

We now discuss the role of anisotropy on the vertex model dynamics. For this purpose, we study the relaxation of the vertex model from an isotropic disordered steady state to an anisotropic steady state (see App.~\ref{sec_vertex_model_implementation} for details of the simulations). 
To understand the transient dynamics between these two steady states, we consider in the following a gradual activation of the anisotropic cell stress or of the anisotropic bond tension, given by:
\begin{align}
    \Sigma^{\rm a }(t) = \Sigma^{\rm a}_0 \left(1-{\rm e}^{-t/T_{\rm a}} \right) \Theta(t) \, , \quad \beta(t)=\beta_0 \left(1-{\rm e}^{-t/T_{\rm a}} \right) \Theta(t) \, ,
\end{align}
where $\Theta(t<0)=0$ and $ \Theta(t\geq0)=1$. We have introduced an activation time $T_{\rm a}$, and $\Sigma^{\rm a}_0$ and $\beta_0$ are the steady-state anisotropic stress magnitude and bond tension magnitude, respectively. 
We have also considered an instantaneous activation where $\Sigma^{\rm a }(t)=\Sigma^{\rm a}_0 \Theta(t)$  and $\beta(t)=\beta_0 \Theta(t)$. This case is presented in App.~\ref{app_instant}.

In the following, we consider vertex model simulations with two types of boundary conditions: (i) \textit{fixed box} boundary condition, for which the box size is fixed and the total tissue shear rate $\widetilde{V}_{ij}$ vanishes; (ii) \textit{stress-free} boundary condition, for which the total stress on the simulation box vanish and $\Sigma_{ij}=0$, such that cells can rearrange and flow. Examples of realization of these two types of boundary conditions are shown in Movies 1 to 4.

    \subsection{T1 transitions driven by anisotropic bond tensions}
    
\begin{figure}[t]
 \centering
    \includegraphics[width=0.48\textwidth]{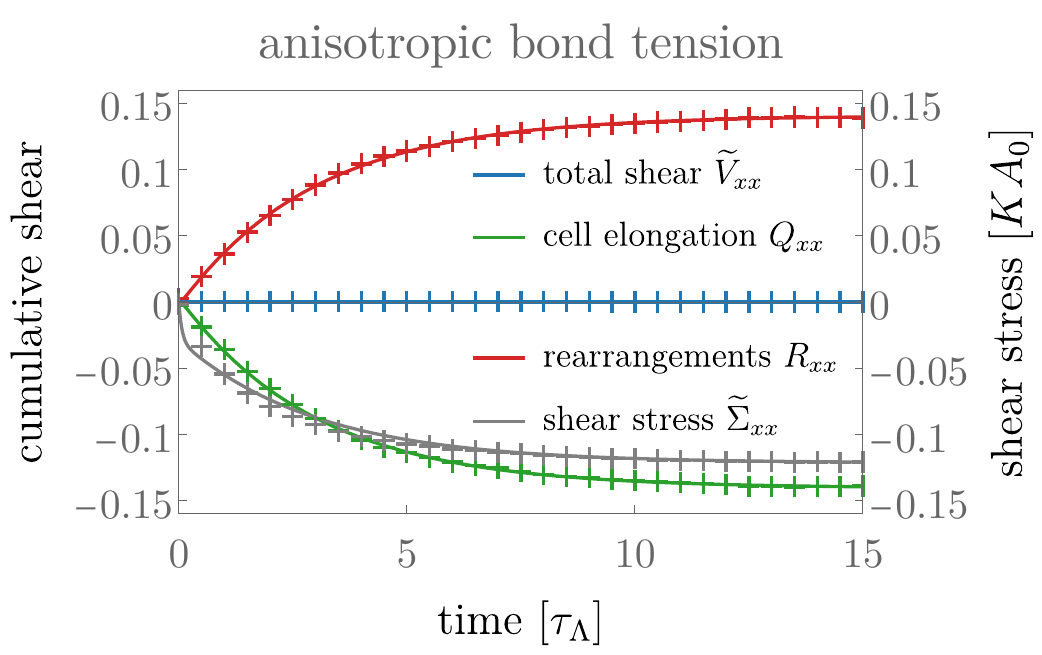}
    \hfill
    \includegraphics[width=0.48\textwidth]{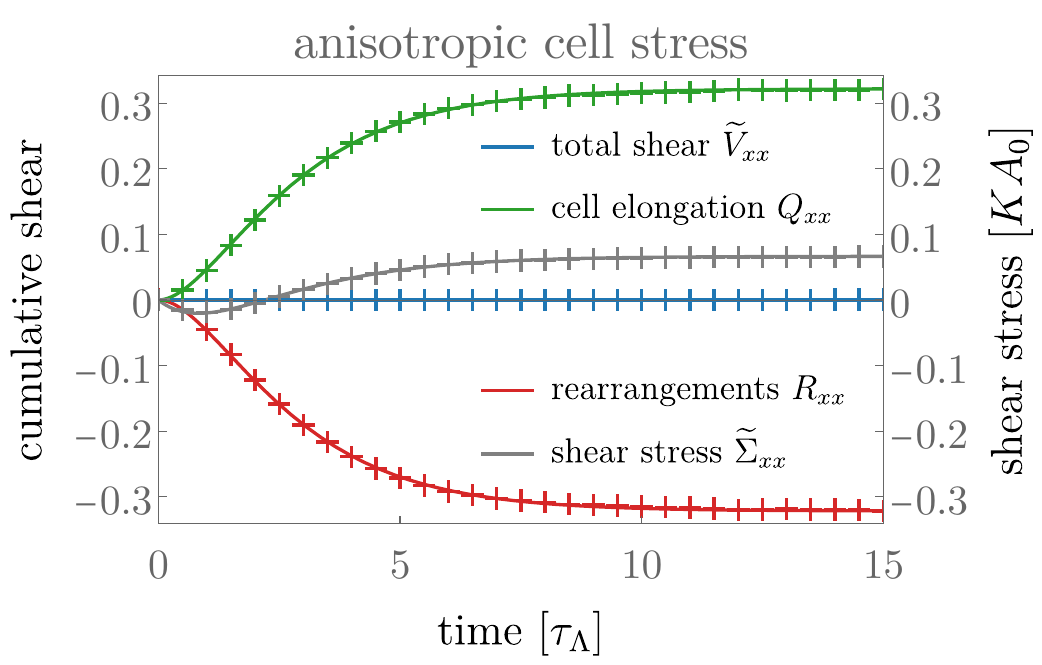}
    
    \vspace{0.5cm}
    
    \includegraphics[width=1\textwidth]{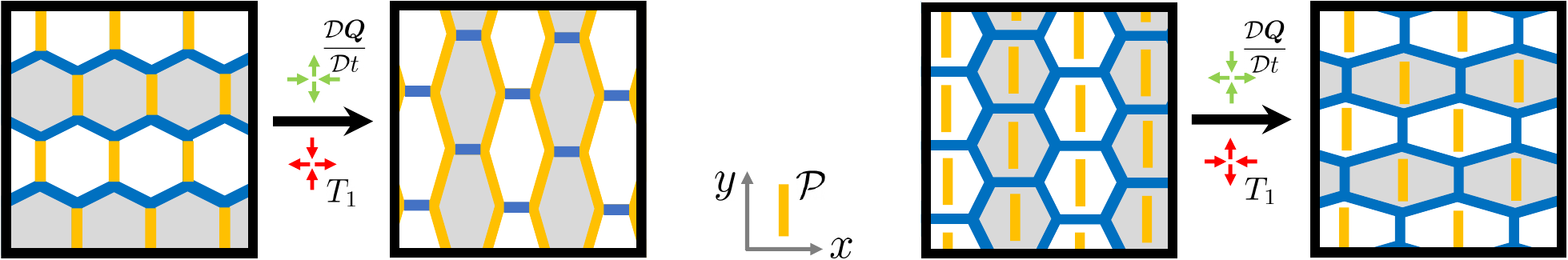}
    
 \caption{%
  Relaxation dynamics after activation of anisotropic active stress under a fixed box boundary condition with anisotropic bond tension~\textbf{(left)} or anisotropic cell stress~\textbf{(right)}. \textbf{Top row.} Total tissue shear (blue) decomposed into contributions of cell elongation change (green) and shear by topological rearrangements (red). The tissue stress is shown in grey.
 Only $xx$-components of the tensors are shown, $xy$-components are zero. Crosses are data from the vertex model averaged over 100 realizations. Error bars are smaller than the marker size. Solid lines are obtained by fits of the continuum model.
 \textbf{Bottom row.} Schematics of the cell rearrangement and elongation explaining the observed dynamics.  
 }
 \label{fig_fixedBox_gradual}
\end{figure}

We now focus on the case of anisotropic bond tension (see Eq.~\eqref{eq_anisotropic_bond_tension}) with a nematic tensor $\bm{\mathcal{P}}$ aligned with the vertical axis, such that bonds are more contractile along this direction. In the continuum model, we translate this choice by taking $p_{xx}=-1$, $p_{yy}=1$ and $p_{xy}=p_{yx}=0$.

The left panel of Fig.~\ref{fig_stressFree_gradual} displays the outcome of vertex model simulations in the case of fixed box boundary conditions. See Movie 1 for an example of a vertex model simulation.
In this case, the larger contractility of bonds along the $y$ axis biases active T1 transitions along the same direction: \new{bonds preferentially close along the $y$ axis and new bonds are opened along the $x$ axis.}
It results in a positive rate of rearrangements along the $x$ axis: $R_{xx}>0$ (red crosses).
This is captured in the continuum model (solid lines) by the fact that the fitted active T1 rate $\lambda$ in Eq.~\eqref{eq_r_constitutive} is positive (see App.~\ref{app_continuumModel} for details of the fitting procedure), meaning that rearrangements are biased in the direction of the nematic tensor.
Additionally, the fixed box imposes that the overall tissue is not sheared ($\widetilde V_{ij}=0$, blue crosses), and cells thus elongate in the direction orthogonal to that of the active T1 (green crosses). See also the lower left panel of Fig.~\ref{fig_fixedBox_gradual} for a schematic explanation of the mechanism.
Note that the tissue stress is along the $y$ direction ($\widetilde\Sigma_{xx}<0$, grey curve). This is consistent with the fact that cells are elongated along the $y$ direction, which induces an elastic stress along the same direction ($\mu q_{xx}<0$ in Eq.~\eqref{eq_sigma_constitutive}). This elastic contribution adds up with the anisotropic one $\zeta p_{xx}$ (with $p_{xx}=-1$), where $\zeta$ is found to be positive from fits to the data, consistent with our definition of anisotropic bond tension which implies a larger stress along the elongation axis.

The case of stress-free boundary conditions is also illuminating, see left panel of Fig.~\ref{fig_stressFree_gradual} and Movie~2. The larger bond tension along the vertical axis leads to shorter vertical bonds, and to an active triggering of T1 transitions that close these short bonds and open horizontal bonds with lower tension (as sketched in the lower left panel of Fig.~\ref{fig_stressFree_gradual}). These active rearrangements (red crosses) drive the shearing of the tissue along the $x$ axis (blue crosses). 
In addition, cells are elongated along the $x$ axis. Indeed, since $\zeta$ is positive, and since $\tilde{\sigma}_{xx}=0$ for stress-free boundary conditions, we deduce from Eq.~\eqref{eq_sigma_constitutive} that $q_{xx}$ is positive.
Note finally that a constant shear rate is obtained in the absence of external driving, which illustrates the active nature of the anisotropic bond tension.

\begin{table}[b]
  \begin{tabular}{|c|c|c|}
    \cline{2-3}
    \multicolumn{1}{c|}{} & \textbf{\, Fixed box \, } &
    \textbf{\, Stress-free \,} \\
    \hline 
    \multirow{2}{2.4cm}{\centering \textbf{anisotropic bond tension}}  & $\dot{\bm{Q}} \parallel \bm{\mathcal{P}}$ &  $\dot{\bm{Q}} \perp \bm{\mathcal{P}}$ \\
    & $\bm{R} \perp \bm{\mathcal{P}}$ & $\bm{R}, \, \widetilde{\bm{V}}  \perp \bm{\mathcal{P}}$\\
    \hline
    \multirow{2}{2.4cm}{\centering \textbf{anisotropic  cell stress}} & $\dot{\bm{Q}} \perp \bm{\mathcal{P}}$ &  $\dot{\bm{Q}} \perp \bm{\mathcal{P}}$ \\
    & $\bm{R} \parallel \bm{\mathcal{P}}$ & $\bm{R}, \, \widetilde{\bm{V}}  \parallel \bm{\mathcal{P}}$\\
    \hline
  \end{tabular}
  \caption{\new{Summary of the steady-state relative orientations. The tensor $\bm{\mathcal{P}}$ gives the direction of the tissue polarity. The tensor $\bm{R}$ indicates the direction of topological transitions (along which new bonds are opened), $\dot{\bm{Q}}=\DDtinline{\bm{Q}}$ is the tensor for the rate of change of cell elongation and $\widetilde{\bm{V}}$ is the tissue shear rate, which indicates the direction along which the tissue elongates.}}
  \label{table_summary_orientation}
\end{table}

    \subsection{T1 transitions driven by anisotropic cell stress}
    
Interestingly, implementing anisotropy via an anisotropic cell stress as defined in Eq.~\eqref{eq_tissue_work_anisotropic} gives rise to a completely different behavior of the cellular network. As in the case of the anisotropic bond tension presented above, we consider a nematic tensor $\bm{\mathcal{P}}$ aligned with the vertical axis, such that cells elongated along the $y$ axis experience a higher stress.

We first consider a fixed box boundary condition, see right panel of Fig.~\ref{fig_fixedBox_gradual} and Movie~3.  
In this case, cells elongate in the direction orthogonal to the nematic axis since the anisotropic cell stress is higher along its direction (green crosses). 
This means, in the continuum model description, that the anisotropic stress magnitude $\zeta$ is positive, as in the case of anisotropic bond tension. 
As a consequence, note that the tissue stress (grey crosses) changes sign during the simulation. At the beginning, tissue stress~\eqref{eq_sigma_constitutive} is dominated by the anisotropic cell stress $\zeta p_{xx}<0$, which is higher in the $y$ direction, leading to a $\tilde\sigma_{xx}<0$. As cells elongate in the $x$ direction in response to this stress, $q_{xx}$ grows and the elastic stress caused by this elongation starts overtaking the anisotropic one and the overall tissue stress changes sign. 
In a fixed box condition, cell elongation has to be compensated by T1 transitions in the opposite direction ($R_{xx}<0$, red crosses), see also bottom right panel of Fig.~\ref{fig_fixedBox_gradual} for a schematic explanation. 
Crucially, \new{the orientation of these T1 transitions (that is, the direction in which new bonds are opened) is orthogonal to the orientation described in the previous section for the anisotropic bond tension (see also Table~\ref{table_summary_orientation}).}
This fact is reflected by the rate $\lambda$ of anisotropic rearrangements in the continuum model, which is now found to be positive for anisotropic cell stress, whereas it was negative for anisotropic bond tension.

\begin{figure}[t]
 \centering
    \includegraphics[width=0.47\textwidth]{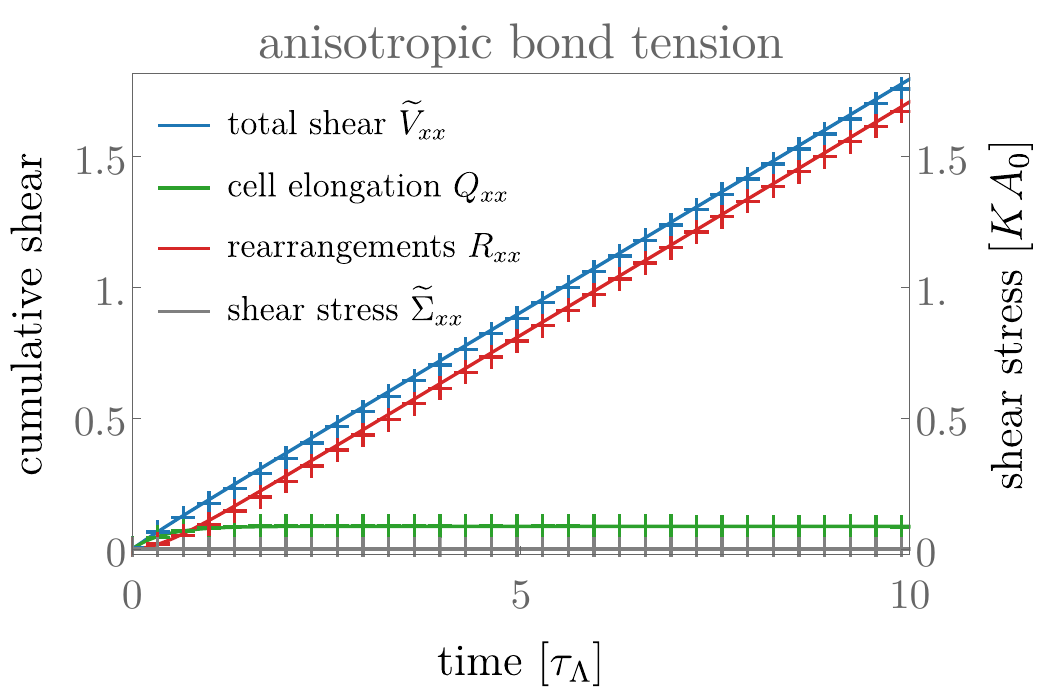}
    \hfill
    \includegraphics[width=0.48\textwidth]{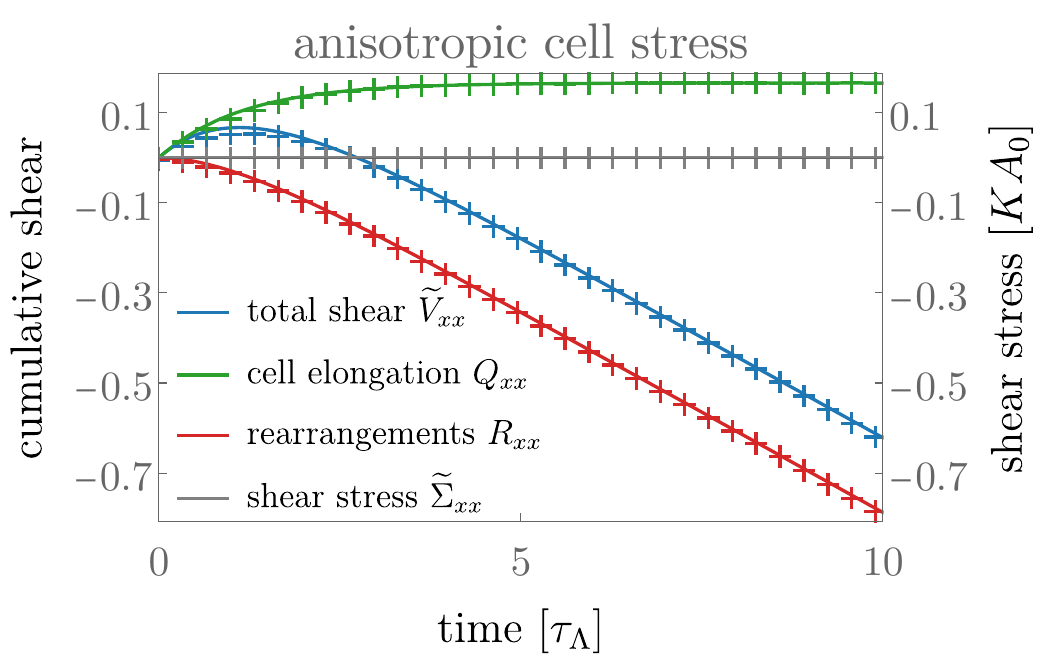}
    
    \includegraphics[width=1\textwidth]{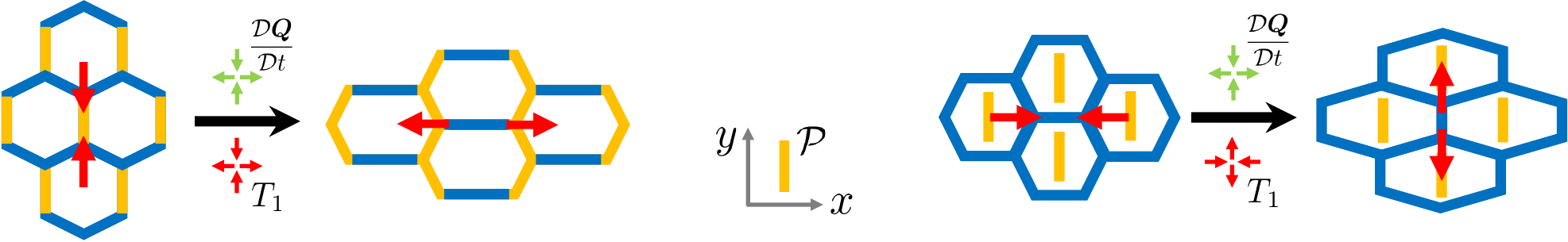}
    
 \caption{%
 Dynamics of tissue shear in a network with anisotropic bond tension \textbf{(left)} or anisotropic cell stress \textbf{(right)} under a stress-free boundary condition. \textbf{Top row.} Total tissue shear (blue) decomposed into contributions of cell elongation change (green) and shear by topological rearrangements (red). The tissue stress is shown in grey (and vanishes as  stress-free boundary conditions are used here).
 Only $xx$-components of the tensors are shown, $xy$-components are zero. Crosses are data from the vertex model averaged over 100 realizations. Error bars are smaller than the marker size. Solid lines are obtained by fits of the continuum model.
 \textbf{Bottom row.} Schematics of the cell rearrangement and elongation explaining the observed dynamics.  
 }
 \label{fig_stressFree_gradual}
\end{figure}

The consequences of an anisotropic cell stress can also be observed in the case of stress-free boundary conditions. Movie~4 shows an example of vertex model simulation in this case, and a quantification in terms of cumulative shear decomposition is displayed on the right panel of Fig.~\ref{fig_stressFree_gradual}.
Anisotropic cell stress drives cells to elongate in the direction orthogonal to the $y$ axis, and we therefore have $Q_{xx}>0$ (green crosses). As a consequence, cells have shorter bonds along their axis of elongation (the $x$ axis) and T1 transitions
\new{close preferentially bonds along this axis and open new ones along the $y$ axis}
%
(hence $R_{xx}<0$, red crosses), see sketch in the lower right panel of Fig.~\ref{fig_stressFree_gradual}. 
The tissue is therefore sheared along the vertical direction (blue crosses), which is opposite to the anisotropic bond tension case. Note also the change of sign of the tissue shear $\widetilde{V}_{xx}$ at short times. At the beginning of the simulation, the anisotropic cell stress immediately drives the elongation of cells, implying $\widetilde{V}_{xx}\simeq \DDtinline{Q_{xx}}>0$ at short time. 
With a delay, T1 transitions respond to this elongation and start contributing to the total tissue shear. They eventually dominate (at $t\gtrsim 3$), and account for the steady-state shear flow.

\new{The different anisotropic stress realizations and boundary conditions lead to different relative steady-state orientations of cell elongation, T1 transitions and tissue shear with respect to the polarity axis. We summarize them in Table~\ref{table_summary_orientation}.}

\section{Energy balance in a tissue}  

In order to define the work performed by T1 transitions, we now discuss the energy balance of a cellular networks subject to active processes and external stresses.
For simplicity and since this work is focused on shear, we limit our discussion to shape changes and shear deformations but do not include changes of tissue size.
In the presence of an external shear stress $\tilde{\bm{\sigma}}^{\rm ext}$ applied to a tissue, the mechanical work per unit time $\dot w_{\rm mech}$ performed on the tissue reads:
\begin{align}
    \dot w_{\rm mech} = \tilde{\bm{\sigma}}^{\rm ext} : \tilde{\bm{v}} \, .
\end{align}
At force balance and for a homogeneous tissue, we have $\tilde{\bm{\sigma}}^{\rm ext}=\tilde{\bm{\sigma}}$. Using the shear decomposition~\eqref{eq_cont_shear_decomp} and the constitutive equations~\eqref{eq_r_constitutive}-\eqref{eq_sigma_constitutive}, the balance of elastic energy $e=\mu \bm{q}:\bm{q}/2$ reads
\begin{align}%
    \dot e &= \dot q_{\rm heat} + \dot w_{\rm mech} + \dot w_{\rm chem}  \, .
\end{align}    
Here, we have defined the power $\dot q_{\rm heat}$ supplied to the system in the form of heat, and the rate of chemical work by the environment on the system $\dot w_{\rm chem}$. These quantities are given by
\begin{align}  
    \dot e &= \mu \bm{q} : \DDt{\bm{q}} \, , \quad
    \dot q_{\rm heat} = -\eta \bm{r}:\bm{r}  \, , \quad
    \dot w_{\rm chem} = \eta \tilde\lambda \bm{p}:\bm{r} - \zeta \bm{p}:\DDt{\bm{q}} \, ,
\end{align}%
where $\eta=\mu\tau$ is the effective tissue viscosity, and we have defined the rate of active T1 transitions $\tilde\lambda=\lambda-\zeta/(\mu\tau)$. Note that the rate of heat production $\dot q_{\rm heat}$ is always negative, indicating that the system releases heat to its surrounding. 

The chemical power $\dot w_{\rm chem}$ is an active contribution that would be vanishing for passive materials. There are two contributions to the chemical power stemming from different processes:
\begin{align}
    \dot w_{\rm chem} = \dot w_{\rm T1} + \dot w_{\rm cell} \, , 
\end{align}
where $\dot w_{\rm T1}=\mu\tau \tilde\lambda \bm{p}:\bm{r}$ is the rate of work by T1 transitions, and $\dot w_{\rm cell}=- \zeta \bm{p}:\DDtinline{\bm{q}}$ is the rate of work by cell deformations. Importantly, these contributions can be either positive or negative. A positive sign indicates that the process is performing work on the tissue, while a negative sign indicates that the process is typically dissipative, but it could also generate chemical free energy. Based on these considerations, we thus define active T1 transitions as T1 transitions for which $\dot w_{\rm T1}>0$.

Determining the effective parameters from vertex model simulations reveal that the rate of active T1 transitions $\tilde\lambda$ is negative for the anisotropic bond tension but positive for anisotropic cell stress (see Table~\ref{table_fit}). However, the chemical work performed by T1 transitions $\dot w_{\rm T1}$ is positive in both cases. This is because  $\bm{p}:\bm{r}$ is negative for anisotropic bond tension, while it is positive for anisotropic cell stress. We thus conclude that T1 transitions are active and perform chemical work on the tissue in both realizations of anisotropic active stress. 
The sign of $\dot w_{\rm T1}$ could become negative if external stress induces shear along an axis perpendicular to the axis of spontaneous shear, i.e., by inducing shear (and rearrangements) along the $y$ axis for the anisotropic bond tension case, or along the $x$ axis in the anisotropic cell stress case. In this case, T1 transitions would be passive and the chemical energy of the active process would be dissipated.

Interestingly, the situation is slightly different for the chemical work performed by cells. Indeed, analysis of the vertex model simulations shows that $\zeta$ is positive both for anisotropic bond tension and anisotropic cell stress (see Table~\ref{table_fit}). For fixed box boundary conditions, $\bm{p}:\DDtinline{\bm{q}}$ is positive for anisotropic bond tension  but it is negative for anisotropic cell stress (see Fig.~\ref{fig_fixedBox_gradual}). This reveals that the work performed by cell deformations $\dot w_{\rm cell}$ is positive and active for anisotropic cell stress, while it is negative and passive for the anisotropic bond tension.
In the case of anisotropic bond tension which drive active T1 transitions, cells elongate for fixed box boundary conditions along the $y$ axis, thus increasing the length of bonds with large contractility, corresponding to $\dot w_{\rm cell}$ being negative and typically dissipative. In contrast, in the case of the anisotropic cell stress, both T1 transitions and cell deformations are active and perform work.
For stress-free boundary conditions, we find that for both anisotropic bond tension and anisotropic cell stress, $\bm{p}:\DDtinline{\bm{q}}<0$ and therefore the work of cell deformations $\dot w_{\rm cell}$ is always positive. Therefore, both T1 transitions and cell deformations are performing work on the tissue to shear it.

\section*{Discussion and conclusion}

Using a vertex model with a preferred axis set by a prescribed nematic field, we have proposed two realizations of anisotropic active processes in tissues. The first one considers anisotropic bond tensions, for which cell bonds aligned with the nematic axis have higher contractility than those oriented perpendicularly. 
The second one involves an anisotropic cell stress aligned with the nematic axis. Importantly, although in both cases an active anisotropic stress exists that is contractile along the nematic axis, these two systems exhibit different orientations of T1 transitions and cell elongation (see Figs.~\ref{fig_fixedBox_gradual} and~\ref{fig_stressFree_gradual}).

In the case of anisotropic bond tension, cell bonds shorten and trigger T1 transitions. For fixed box boundary conditions, we therefore observe that cells elongate along an axis parallel to the nematic axis and orthogonal to the \new{orientation of bonds opened by} T1 transitions. For stress-free boundary conditions, both cell elongation and T1 transitions are perpendicular to the nematic axis.
Anisotropic cell bond tension captures the behavior observed during the early stages of pupal wing development in \textit{Drosophila}~\cite{etournay2015interplay}. In that case, the increased contractility is oriented along the proximal-distal axis. The phenomenological parameters were measured as $\zeta/\mu\simeq0.33$, $\tau\simeq1.7$~h, and $\lambda\simeq-0.11$~h$^{-1}$, which corresponds to $\tilde\lambda\simeq-0.31$~h$^{-1}$. This suggests that T1 transitions are active in these early stages and driven by anisotropic bond tension.
Overall tissue shear was smaller than the cell shape change, corresponding to a case where the boundaries are slowly moving, not too far from the fixed box boundary conditions.

A different situation arises in the case of anisotropic cell stress, in which cells elongate and trigger T1 transitions.
For fixed box boundary conditions, cells elongate perpendicular to the nematic axis, and T1 transitions are oriented parallel to the nematic axis. These orientations remain the same for stress-free boundary conditions (see Figs.~\ref{fig_fixedBox_gradual} and~\ref{fig_stressFree_gradual}).
\new{Anisotropic cell stress could contribute to} the behavior observed during germ-band extension in the \textit{Drosophila} embryo~\cite{bertet2004myosindependent,collinet2015local}. The tissue extends along the anterior-posterior (AP) axis, which suggests that anisotropic active stress is orthogonal to this axis. Both in wild type and when tissue extension is obstructed by laser cauterization, cells elongate perpendicular to the AP axis. These observations are consistent with anisotropic cell stress both for fixed box boundary conditions (cauterization) and stress-free boundary conditions (rough approximation for wild type). In addition, it was reported that tissue elongation was driven by anisotropic medial myosin II pulses~\cite{collinet2015local}, which are expected to generate anisotropic cell stress.

We have considered these two realizations of anisotropic active processes \textit{separately}. However, in biological tissues, both types of active stresses could coexist.
\new{This is likely the case during \textit{Drosophila} germ-band extension, where an anisotropic accumulation of myosin II at cell junctions is observed~\cite{rauzi2008nature,tetley2016unipolar,wang2020anisotropy} as well as anisotropic pulses of medial actin~\cite{collinet2015local}, indicating the existence of anisotropic cell stress. This suggests that both types of active anisotropic processes could be simultaneously relevant.}

\new{If both processes are at work at the same time with the polarity $\bm{\mathcal{P}}$ aligned to the same axis, they would be antagonistic and oppose each other. This case is similar to a tug-of-war situation where the resulting T1 transitions would be the net result of the two opposing anisotropic processes generating shear along orthogonal axes.}
One can speculate that the relative strength of these two opposing processes could be fine-tuned such that the resulting net rate of active T1 transitions would be vanishing, even though the system would still be chemically active and T1 transitions fluctuate strongly forward and backward. This could lead to a fluidization of the tissue \new{or give rise to stress oscillations}. Furthermore, at this balance point, a biological tissue could be capable of changing rapidly to one of the two steady states with orthogonal shear axis if the balance between the opposing activities is lost.
\new{Recent work has proposed more detailed descriptions of T1 transitions, including delay~\cite{erdemci-tandogan2021effect} or nonlinear dynamics~\cite{krajnc2021active}. 
While we expect the qualitative picture studied in this manuscript to be rather robust, it will be interesting to study the role of such nonlinear or delayed T1 transitions in the anisotropic active processes discussed here.}

Using a linear continuum description that captures the anisotropic dynamics of the vertex model, we have shown that the difference between these two realizations of active stress is captured by a relative sign difference between the active stress magnitude $\zeta$ (positive in both scenarios) and the active T1 rate $\lambda$ (positive for anisotropic cell stress, negative for anisotropic bond tension).
Despite these differences, an analysis of the energy balance in the system reveal that T1 transitions perform chemical work on the tissue in both cases, and can therefore be referred to as active.
The determination of~$\zeta$ and~$\lambda$ experimentally is a challenge. However, the determination of the rate of active T1 transition $\tilde\lambda=\lambda-\zeta/\mu\tau$ may be accessible by state-of-the-art experimental techniques. Indeed, at steady state and for fixed box boundary conditions, the tissue shear stress reads $\bm{\tilde{\sigma}}=-\mu\tau\tilde\lambda\bm{p}$ and could therefore provide a readout for the sign of this activity coefficient. This local tissue stress could for instance be measured by injecting liquid oil droplets, as recently shown in the zebrafish embryo~\cite{mongera2018fluidtosolid,kim2021embryonic}.

\subsection*{Acknowledgements}

C.D. thanks Aboutaleb Amiri and Marko Popović for stimulating discussions. 
M.M.I. acknowledges funding from the Indian Science and Engineering Research Board (MTR/2020/000605) and the hospitality at MPI-PKS, Dresden. 
C.D.M and F.J. acknowledge funding by the German Federal Ministry of Education and Research under grant number 031L0160.

\subsection*{Author contribution statement}

All authors contributed equally to the paper.

\newpage

\appendix

\input{appendix}

\bibliographystyle{apsrev4-1}
\bibliography{biblio.bib}

\end{document}

%% file: appendix.tex
\section{Vertex model simulations}
\label{sec_vertex_model_implementation}
In this appendix, we provide a more detailed description of the vertex model simulations.

\begin{table}[b]
  \begin{tabular}{ |c|l|r|c| }
    \hline
    \multicolumn{4}{|c|}{vertex model parameters} \\
    \hline
    \textbf{symbol} & \textbf{explanation} & \textbf{value} & \textbf{unit} \\
    \hline
    \multicolumn{4}{|c|}{mechanics} \\
    \hline
    $\Lambda_0$         & mean line tension    & 0.12 & $K A_0^{3/2}$ \\ 
    $\Gamma$            & perimeter elasticity & 0.04 & $K A_0$ \\ 
    $\beta_0$            & steady-state line tension anisotropy & 0.50 & - \\ 
    $\Sigma^\mathrm{a}_0$ & steady-state active cell stress magnitude & 0.04 & $K A_0$ \\ 
    $T_\mathrm{a}$ & anisotropy activation time & 1 & $\tau_\Lambda$ \\ 
    \hline
    \multicolumn{4}{|c|}{dynamics} \\
    \hline
    $\Delta\Lambda$     & line tension fluctuations magnitude & 0.06 & $K A_0^{3/2}$ \\ 
    $\delta t$          & time step for numerical integration & 0.01 & $\tau_\Lambda$ \\ 
    \hline
  \end{tabular}
  \caption{Parameter values used in the simulations of the vertex model, expressed in dimensionless units. 
  In case different values are used for simulation results, it is stated in the caption of the figure showing the results.}
  \label{tab_vertex_model_parameters}
\end{table}

\subsection{Dimensionless parameters}
In our simulations of the vertex model we use parameters expressed in dimensionless units.
To this end, we will choose $\tau_\Lambda$ as the typical timescale, $A_0^{1/2}$ as the typical length scale and $K A_0^2$ as the typical energy scale in our model.
The effective parameters are thus reduced to \mbox{$\bar{\Lambda}_{mn}=\Lambda_0/(K A_0^{3/2})$}, \mbox{$\bar{\Gamma}=\Gamma/(K A_0)$}, \mbox{$\Delta\bar{\Lambda}=\Delta\Lambda/(K A_0^{3/2})$}.
The parameters $\tau_\Lambda$, $K$, and $A_0$ are unity in these units.
Throughout the manuscript, we use dimensionless units and omit the bar on the parameter symbols for simplicity.
We choose the parameter values $\Lambda_0=0.12$ and $\Gamma=0.04$ which are known to produce network configurations that agree well with those observed in the wing disk epithelium \cite{farhadifar2007influence}.
Other parameter values are listed in Table~\ref{tab_vertex_model_parameters}.

\subsection{Model initialization and simulation}

Each simulation is initialized as a network of $N_x=20$ by $N_y=20$ regular hexagonal cells, with box dimensions $L_x$ and $L_y$ set such that the network work function~\eqref{eq_tissue_work} is in its ground state~\cite{farhadifar2007influence}. 
We first propagate the system under a fixed boundary condition and without anisotropy ($\beta_0=\Sigma^{\rm a}_0=0$), until the system has reached a steady state configuration at $t=50\tau_\Lambda$. Once the system is prepared, we distinguish the following two cases.

(i) \textit{Fixed box boundary condition.} In this case, after reaching an isotropic steady state, the fixed box boundary condition is kept and one of the two anisotropic contributions is included. In the case of an instantaneous activation, this contribution is included in the work function immediately at the end of the equilibration time. In the case of a gradual activation, the anisotropic contribution is added with an exponential increase $1-{\rm e}^{-t/T_{\rm a}}$. The system is then propagated for a time $t=15\tau_\Lambda$. This procedure is repeated $N=100$ times and the corresponding simulations results are displayed in Fig.~\ref{fig_fixedBox_gradual} for the gradual activation case and in Fig.~\ref{fig_fixedBox_instant} for the instantaneous activation case.

(ii) \textit{Stress-free boundary condition.} In this case, after reaching an isotropic steady state under a fixed box boundary condition, one of the two anisotropic contributions is included and the box is left free to deform such that no stress is exerted on the simulation box. Similarly to the fixed box case, gradual and instantaneous activations are considered. The system is propagated for a time $t=10\tau_\Lambda$. This procedure is repeated $N=100$ times and the corresponding simulations results are displayed in Fig.~\ref{fig_stressFree_gradual} for the gradual activation case and in Fig.~\ref{fig_stressFree_instant} for the instantaneous activation case.

\subsection{T1 transitions}

In this implementation of the vertex model, a full T1 transition is divided into 2 steps. First, two connected three-fold vertices that reach a distance $\mathcal{L}_{mn}$ lower than a certain threshold value, will merge to form a four-fold vertex. Next, a tentative split into two vertices is attempted in both possible topologies of the network. 
If in one of the topologies the two vertices are pulled apart by a force, the four-fold vertex is unstable. If the four-fold vertex is unstable for both topologies, the topology that maximizes the force magnitude is chosen. In the case where the four-fold vertex is stable, it is kept as such and a new tentative split is attempted at the next time step.
In case that cell neighbors have changed as compared to before the merger of the vertices, we call it a full T1 transition. Upon a T1 transition, which creates of a new bond at time $t_0$, the initial value of the bond tension $\Lambda_{mn}(t_0)$ is drawn from a normal distribution with mean $\bar{\Lambda}_{mn}$ and variance $\Delta\Lambda$. After the new bond is created, the system is again relaxed to a force-balanced state.

\subsection{Constant cell number ensemble}

\label{sec_constant_cell}

In all the vertex model simulations in this paper, we use a fixed cell number ensemble. In some rare cases, bond tension fluctuations can drive the area of a cell below a critical area. Below this critical area, the cell area imposed by the local minimum of this cell's work function is zero. The cell would therefore shrink to have an area that is zero. When the cell area reaches a value below a certain set threshold, the cell is extruded from the network and replaced by a vertex which has the same order as the neighbor number of the removed cell.
In order to keep the number of cells in the tissue constant, a randomly chosen cell divides.

In case a cell divides, a new bond is created running through the cell center. The new bond makes an angle with the $x$-axis which is drawn from a uniform distribution between 0 and $\pi$. The two daughter cells have the same preferred area as the mother cell, simplifying the more realistic case of the continuous growth of the cell area. 
After each cell division, the configuration of the cell network is changed in order to minimize the work function.

\section{Notation and definitions}
\label{sec_definitions}

        \subsection{Velocity gradient tensor}
        
The motion of cells in the tissue is described by the coarse-grained cell velocity field $v_j$ (or $V_j$ for the vertex model). Deformations of the network are proportional to gradients in this velocity field $v_{ij} = \partial_i v_j$, where $v_{ij}$ is the velocity gradient tensor. The trace of this tensor, $v_{kk}$ (summation over repeated Cartesian indices is implied), corresponds to local isotropic growth of the tissue. The traceless-symmetric part of the velocity gradient tensor, denoted $\tilde v_{ij}$, corresponds to anisotropic deformations, and its antisymmetric part, the vorticity tensor $\omega_{ij}$, characterizes local rotations. The velocity gradient tensor can thus be decomposed as:
\begin{equation}
 v_{ij} = \frac{1}{2} v_{kk} \delta_{ij} + \tilde v_{ij} + \omega_{ij},
 \label{eq_cont_velgrad_decomp}
\end{equation}
where in two dimensions we have $\omega_{ij}=-\omega\varepsilon_{ij}$ with $\varepsilon_{ij}$ the generator of counterclockwise rotation with $\varepsilon_{xy} = -1$, $\varepsilon_{yx}=1$ and $\varepsilon_{xx}=\varepsilon_{yy}=0$.

        \subsection{Stress tensor}

Similarly to the velocity gradient tensor, the tissue stress tensor $\sigma_{ij}$ can also be decomposed into
\begin{align}
    \sigma_{ij} = \frac{\sigma_{kk}}{d}\delta_{ij} + \tilde \sigma_{ij} \, .
\end{align}
where $d$ is the spatial dimension and $\tilde \sigma_{ij}$ is the shear stress. Note that in the absence of chiral terms the stress tensor is symmetric, and we have therefore not included the  antisymmetric contribution to the previous equation.

        \subsection{Corotational time derivative of tensors}

The corotational time derivative of a tensor $M_{ij}$ is defined as:
\begin{equation}
 \DDt{M_{ij}} = \ddt{M_{ij}} + \omega_{ik} M_{kj} + \omega_{jl} M_{il}.
 \label{eq_SI_hydr_model_corotational_derivative_simple_shear}
\end{equation}
where $\omega_{ij}$ is the vorticity tensor of the fluid.

        \subsection{Nematic tensors in two dimensions}
    \label{sec_nematic_tensor_decomposition}
    
In dimension two, a traceless symmetric tensor $M_{ij}$ (that we call nematic tensor) has two degrees of freedom and can be written in terms of its Cartesian coordinates as:
\begin{align}
    \bm{M} = 
    \begin{pmatrix}
    M_{xx} & M_{xy}  \\
    M_{xy} & -M_{xx}
    \end{pmatrix} \, ,
\end{align}
or it can equivalently be decomposed into a norm $M=|\bm{M}|$ and angle $\Theta$ as:
\begin{align}
    \bm{M} = M
    \begin{pmatrix}
    \cos(2\Theta) & \sin(2\Theta)  \\
    \sin(2\Theta) & -\cos(2\Theta) 
    \end{pmatrix} \, , \label{eq_definition_nematic_2d}
\end{align}
with $M=\sqrt{M_{xx}^2 + M_{xy}^2}$ and $\Theta=\frac{1}{2} \arctan (M_{xy},M_{xx})$, where the function $\arctan$ gives the arc tangent of $M_{xy}/M_{xx}$, taking into account in which quadrant the point $(M_{xy},\, M_{xx})$ lies. From Eq.~\eqref{eq_definition_nematic_2d}, one directly sees that $\bm M^2=M^2\mathbb{1}$ for a nematic tensor in two dimensions.

\section{Instantaneous activation of the anisotropy}
\label{app_instant}

\begin{figure}[t]
 \centering
    \includegraphics[width=0.48\textwidth]{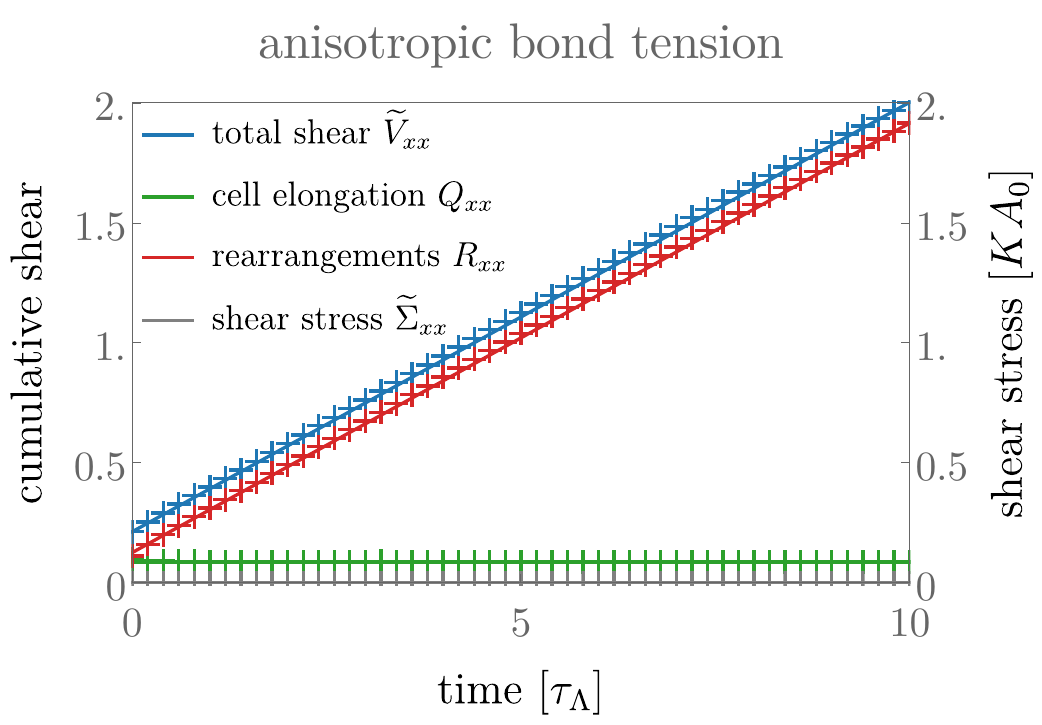}
    \hfill
    \includegraphics[width=0.48\textwidth]{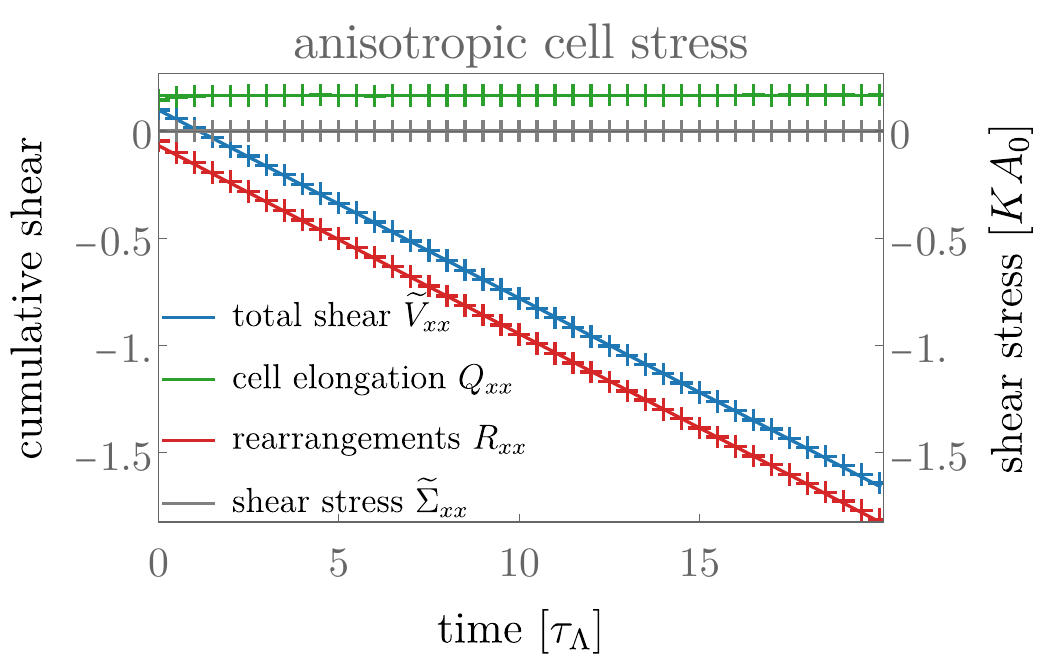}
 \caption{%
 Anisotropic vertex model simulations with stress-free boundary conditions and instantaneous activation of the anisotropic bond tension (left) or anisotropic cell stress (right). Crosses are obtained by averaging 100 realizations of the vertex model simulations (error bars are smaller than the marker size); solid lines are obtained by fitting the hydrodynamic model (see App.~\ref{app_continuumModel} for details and values of the fitted parameters).
 }
 \label{fig_stressFree_instant}
\end{figure}

\begin{figure}[t]
 \centering
    \includegraphics[width=0.48\textwidth]{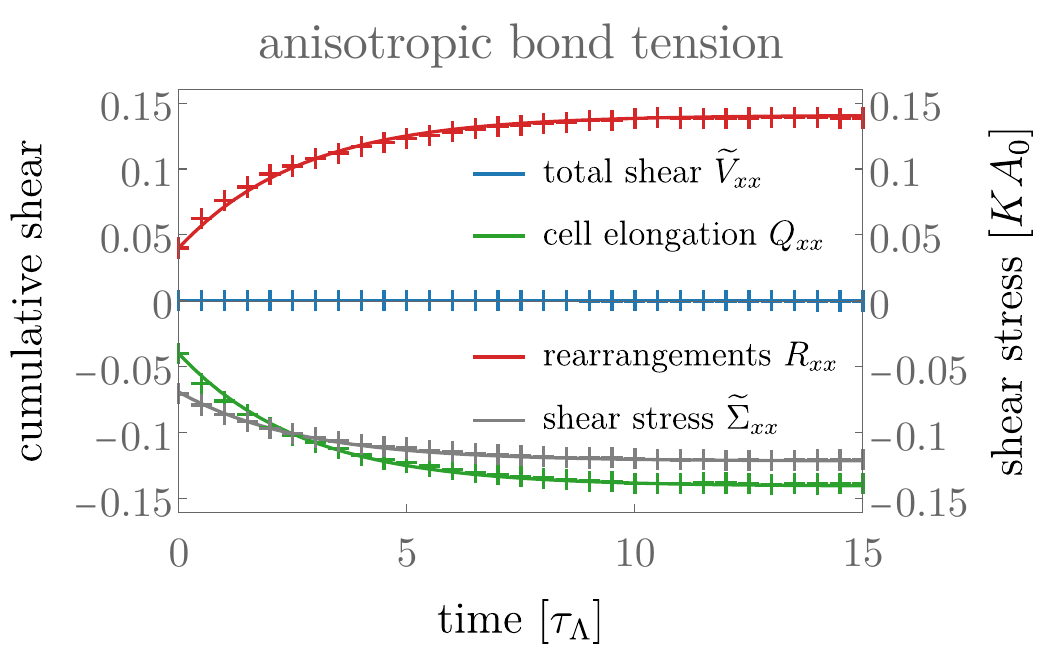}
    \hfill
    \includegraphics[width=0.48\textwidth]{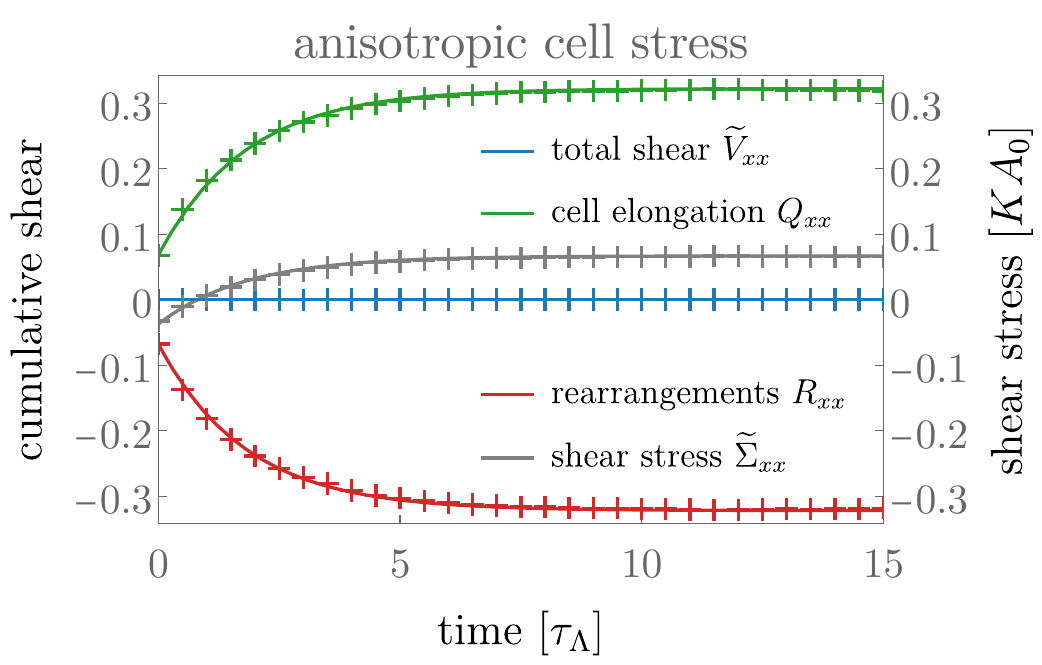}
 \caption{%
 Anisotropic vertex model simulations with fixed box boundary conditions and instantaneous activation of the anisotropic bond tension (left) or anisotropic cell stress (right). Crosses are obtained by averaging 100 realizations of the vertex model simulations (error bars are smaller than the marker size); solid lines are obtained by fitting the hydrodynamic model (see App.~\ref{app_continuumModel} for details and values of the fitted parameters).
 }
 \label{fig_fixedBox_instant}
\end{figure}

In addition to the gradual exponential activation of the anisotropy that we have discussed in the main text, we have also considered an instantaneous activation. In this case, after equilibrating the system with isotropic properties for a time $t=50\tau_\Lambda$, the anisotropic cell stress or the anisotropic bond tension are immediately set to their steady-state values. An exponential relaxation of cell elongation to its steady-state value is observed and fitted to the continuum model (see App.~\ref{app_continuumModel} for details on the fitting procedure). We display in Fig.~\ref{fig_stressFree_instant} the results of the simulations and fits in the case of stress-free boundary conditions, and in Fig.~\ref{fig_fixedBox_instant} the case of a fixed box boundary condition.

\section{Continuum model for anisotropic tissues}
\label{app_continuumModel}

Here, we discuss how the continuum model is used to fit the vertex model data. We recall for convenience the shear decomposition and the constitutive equations for the tissue stress and rate of cellular rearrangements:
\begin{subequations}%
\begin{align}%
    \tilde v_{ij} &= \DDt{q_{ij}} + r_{ij} \, , \label{SI_eq_shear_decomp} \\
    r_{ij} &= \frac{1}{\tau} q_{ij} + \lambda(t) p_{ij} \, ,  \label{SI_eq_r_constitutive} \\
    \tilde{\sigma}_{ij} &= \mu q_{ij} + \zeta(t) p_{ij} \, . \label{SI_eq_stress_constitutive}
\end{align}%
\end{subequations}%
The dynamics and steady states depend crucially on the imposed boundary conditions, and we discuss below the stress-free and fixed box boundary conditions studied in the main text.

    \subsection{Fixed box boundary condition}

For a fixed box boundary condition, the tissue cannot deform and $\tilde v_{ij}=0$, such that shear decomposition~\eqref{SI_eq_shear_decomp} reduces to:
\begin{align}
    q_{ij}'(t) = - r_{ij} (t) \, , \label{eq_SI_fixedBox_shearDecomp}
\end{align}
where the prime denotes the time derivative.

        \subsubsection{Steady state}
        
At steady state, cell elongation is constant, and we deduce from Eq.~\eqref{eq_SI_fixedBox_shearDecomp}  that $r_{ij}=0$. We thus obtain:
\begin{align}
    q_{ij}^{\rm ss} = - \lambda_0\tau p_{ij} \, , \quad \tilde\sigma_{ij}^{\rm ss} = \left(-\mu\tau\lambda_0 +\zeta_0 \right) p_{ij} = -\mu\tau \tilde\lambda_0 p_{ij} \, ,
\end{align}
where $\lambda_0$ and $\zeta_0$ are the steady-state values of $\lambda(t)$ and $\zeta(t)$, respectively.   We have also introduced $\tilde\lambda_0=\lambda_0-\zeta_0/(\mu\tau)$. The coefficients $-\lambda_0\tau$ and $-\mu\tau\tilde\lambda_0$ are then obtained from the vertex model data by computing the steady-state mean cell elongation and the steady-state tissue shear stress, respectively.

Note that in the main text we only discuss the steady-states values $\lambda_0$, $\zeta_0$, and $\tilde \lambda_0$ of the parameters, and we have dropped their subscript $0$ for simplicity.
        
        \subsubsection{Instantaneous activation}
        
We consider instantaneous activation of the anisotropy in the vertex model. We therefore consider $\lambda(t)=\lambda_0 \Theta(t)$ and $\zeta(t)=\zeta_0 \Theta(t)$  where $\Theta(t<0)=0$ and $\Theta(t\geq0)=1$ in the continuum description.      
In this case, Eq.~\eqref{SI_eq_shear_decomp} can be solved and yields:
\begin{align}
    q_{ij} (t) = q_{ij}(t=0^+){\rm e}^{-t/\tau}-\lambda_0 \tau\left( 1-{\rm e}^{- t/\tau} \right) p_{ij} \, ,
    \label{eq_SI_fixedBox_instant}
\end{align}
where $q_{ij}(t=0^+)$ is the value of cell elongation immediately after activation of the anisotropy. The exponential relaxation given by Eq.~\eqref{eq_SI_fixedBox_instant} can be fitted to the vertex model data for cell elongation to obtain the timescale $\tau$. 

From this fit and the values obtained from steady state, the only remaining parameter is $\mu$, which is obtained from fitting the continuum model shear stress exponential relaxation
\begin{align}
    \tilde\sigma_{ij} (t) = \mu q_{ij}(t=0^+){\rm e}^{-t/\tau} + \left[ \zeta_0-\lambda_0\mu \tau\left( 1-{\rm e}^{-t/\tau} \right) \right] p_{ij} \, ,
\end{align}
to the tissue shear stress from the vertex model data.

        \subsubsection{Gradual activation}

We consider an exponential gradual adaptation of the anisotropy in the vertex model. In the continuum description, we introduce two adaptation times $\tau_\lambda$ and $\tau_\zeta$, such that $\lambda(t)=\lambda_0(1-{\rm e}^{-t/\tau_\lambda})$ and $\zeta(t)=\zeta_0 (1-{\rm e}^{-t/\tau_\zeta})$.
Solving Eq.~\eqref{SI_eq_shear_decomp} with $q_{ij}(0)=0$, we obtain:
\begin{align}
    q_{ij} (t) = -\lambda_0 \tau p_{ij}\left(1 + \frac{ \tau {\rm e}^{-t/\tau} + \tau_\lambda{\rm e}^{-t/\tau_\lambda}}{\tau- \tau_\lambda} \right) \, .
    \label{eq_SI_fixedBox_gradual}
\end{align}
This solution is fitted against the data from the vertex model for cell elongation to obtain~$\tau_\lambda$ and~$\tau$.
Using these fitted values and the parameter values obtained from steady state, the remaining parameters to obtain are~$\mu$ and~$\tau_\zeta$. They are obtained from fitting the continuum model shear stress
\begin{align}
    \tilde\sigma_{ij} (t) = \left[ \zeta_0\left( 1 - {\rm e}^{-t/\tau_\zeta} \right)
    -\lambda_0\mu \tau \left(
    1 + \frac{ \tau {\rm e}^{-t/\tau} + \tau_\lambda{\rm e}^{-t/\tau_\lambda}}{\tau- \tau_\lambda}
    \right)\right] p_{ij} \, ,
\end{align}
to the tissue shear stress from the vertex model simulations.
Note that the fits displayed in Figs.~\ref{fig_fixedBox_gradual} and~\ref{fig_stressFree_gradual} of the main text have been obtained using a single activation timescale $\tau_{\rm a}=\tau_\lambda=\tau_\zeta$. 

    \subsection{Stress-free boundary condition}

Under a stress-free boundary condition, we impose the total stress $\sigma_{ij}$ to vanish. The tissue is left free to deform and the tissue deformation tensor $v_{ij}$ is unconstrained. A vanishing stress implies directly from Eq.~\eqref{SI_eq_stress_constitutive} that cell elongation and the anisotropic stress are proportional:
\begin{align}
    q_{ij} (t) = - \frac{\zeta (t)}{\mu} p_{ij} \, .\label{eq_SI_stressFree_stress}
\end{align}

        \subsubsection{Steady state}
        
At steady state, we obtain for stress-free boundary conditions:
\begin{align}
    q_{ij}^{\rm ss} = -\frac{\zeta_0}{\mu}p_{ij} \, , \quad \tilde v_{ij}^{\rm ss} =
    \tilde\lambda_0 p_{ij} \, ,
\end{align}
where $\tilde\lambda_0=\lambda_0-\zeta_0/(\mu\tau)$ and where $\zeta_0$ are the steady-state values of $\lambda(t)$ and $\zeta(t)$, respectively. The coefficients $-\zeta_0/\mu$ and $\tilde\lambda_0$ are then obtained from the vertex model data by computing the steady-state mean cell elongation and the steady-state tissue shear rate, respectively.     

Note that in the main text we only discuss the steady-states values $\lambda_0$, $\zeta_0$, and $\tilde \lambda_0$ of the parameters, and we have dropped their subscript $0$ for simplicity.
        
        \subsubsection{Instantaneous activation}
        
We consider instantaneous activation of the anisotropy in the vertex model. We therefore consider $\lambda(t)=\lambda_0 \Theta(t)$ and $\zeta(t)=\zeta_0 \Theta(t)$  where $\Theta(t<0)=0$ and $\Theta(t\geq0)=1$ in the continuum description. In this case and for stress-free boundary conditions, we remark that Eq.~\eqref{eq_SI_stressFree_stress} implies $\tilde v_{ij}= r_{ij}$, and the instant adaptation case can be fitted using only the parameter values extracted from the steady-state analysis.

        \subsubsection{Gradual activation}

We consider an exponential gradual adaptation of the anisotropy in the vertex model. In the continuum description, we introduce two adaptation times $\tau_\lambda$ and $\tau_\zeta$, such that $\lambda(t)=\lambda_0(1-{\rm e}^{-t/\tau_\lambda})$ and $\zeta(t)=\zeta_0 (1-{\rm e}^{-t/\tau_\zeta})$. In this case,
Eq.~\eqref{SI_eq_stress_constitutive} reads:
\begin{align}
    q_{ij} (t) = -\frac{\zeta_0}{\mu} \left(1-{\rm e}^{-t/\tau_\zeta}  \right) p_{ij} \, .
\end{align}
This equation can be fitted to the cell elongation data to obtain the timescale $\tau_\zeta$. The second activation timescale~$\tau_\lambda$ as well as $\tau$ can then be obtained by fitting the continuum model shear rate
\begin{align}
    v_{ij}(t) = \left[\tilde \lambda_0 - \left( \frac{\zeta_0}{\mu\tau} + \tilde \lambda_0 \right) {\rm e}^{-t/\tau_\lambda}
    + \frac{\zeta_0}{\mu} \left( \frac{1}{\tau} - \frac{1}{\tau_\zeta} \right) {\rm e}^{-t/\tau_\zeta} \right] p_{ij}
\end{align}
to the vertex model shear rate. Note that the fits displayed in Figs.~\ref{fig_fixedBox_gradual} and~\ref{fig_stressFree_gradual} of the main text have been obtained using a single activation timescale $\tau_{\rm a}=\tau_\lambda=\tau_\zeta$.

    \subsection{Parameter values obtained from the fits}
    
In Table~\ref{table_fit}, we display the parameters obtained from fitting the continuum model to the vertex model simulations. Note that for the stress-free boundary condition, only the reduced parameters $\zeta_0/\mu$ and $\tilde\lambda_0=\lambda_0-\zeta_0/(\mu\tau)$ can be obtained.   

For the anisotropic cell stress case, the agreement between the four realizations of the model (fixed box or stress-free boundary conditions, and gradual or instantaneous activation) is excellent, and the discrepancy between fitted values is minimal.
For the anisotropic bond tension case, we however observe relatively strong discrepancies between the different fits between the different activation procedures and boundary conditions.
This is likely due to nonlinear properties of the vertex models, not captured by our linear version. 
Indeed, these nonlinear properties have been shown to occur, for the same vertex model parameters as those studied here, at relatively large shear rate of order $\tilde{v}_{xx}\sim0.2$~\cite{duclut2021nonlinear}. In the anisotropic model that we consider here, the system is not sheared by an externally-imposed shear rate, but the shearing is rather induced by the anisotropic activity. In the stress-free boundary condition (see Figs.~\ref{fig_stressFree_gradual} and~\ref{fig_stressFree_instant}), we can compute this induced shear rate and find that it is $\tilde{v}_{xx}\simeq 0.18$ for our choice of parameters and for anisotropic bond tension, and $\tilde{v}_{xx}\simeq -0.089$ for anisotropic cell stress. The anisotropic cell stress version is therefore driven at a lower shear rate and is thus better captured by the linear model. 
A nonlinear version of the model, constructed in the spirit of what has been done in Ref.~\cite{duclut2021nonlinear}, is beyond the scope of this work.

\begin{table}[t]
  \begin{tabular}{|c|c||c|c|c|c|c|c|c||c|c|c| }
    \cline{3-12}
    \multicolumn{2}{c}{}  & \multicolumn{10}{|c|}{Parameter values obtained from fit} \\
    \cline{3-12}
    \multicolumn{2}{c}{} & \multicolumn{7}{|c||}{\textbf{Fixed box}} & \multicolumn{3}{c|}{\textbf{Stress-free}} \\
    \cline{3-12}
    \multicolumn{2}{c|}{} & $\mu $ & $\tau $ & $\lambda_0$ & $\zeta_0$ & $\zeta_0/\mu$ & $\tilde\lambda_0$ & $\tau_\mathrm{a}$ & $\zeta_0/\mu$ & $\tilde\lambda_0$ & $\tau_\mathrm{a}$ \\
    \hline \hline
    \multirow{2}{2.4cm}{\centering \textbf{anisotropic bond tension}}  & gradual activation & \, 0.64 \, & \,3.0\, &  \,-0.047\, & \,0.031\, & \,0.048\, & \,-0.063\, & \,0.068\, & \,0.087\, & \,-0.18\, & \,0.41\, \\
    & instant. activation& 0.52 & 2.7 &  -0.052 & 0.048 & 0.093 & -0.087 & / & 0.087 &  -0.18 & / \\
    \hline
    \multirow{2}{2.4cm}{\centering \textbf{anisotropic  cell stress}} & gradual activation & 0.42 & 1.9 & 0.17 & 0.069 & 0.16 & 0.085 & 1.2 & 0.17 & 0.089 & 1.2\\
    & instant. activation & 0.41 & 1.8 &  0.18 & 0.064 & 0.16 & 0.090 & / & 0.17 &  0.088 &  / \\
    \hline
  \end{tabular}
  \caption{
  Parameter values for the hydrodynamic model with anisotropy, obtained by fitting Eqs.~\eqref{SI_eq_shear_decomp} to~\eqref{SI_eq_stress_constitutive} to the vertex model simulation data. Details of the fitting procedure can be found in App.~\ref{app_continuumModel}. Note that in the main text we have dropped the subscript $0$ of $\lambda_0$, $\zeta_0$, and $\tilde \lambda_0$ for simplicity.}
  \label{table_fit}
\end{table}

\section{Supplemental movies}
\label{sec_movies}

\subsection{Movie 1: anisotropic bond tension with fixed boundary conditions}

\noindent\textbf{Description:} Snapshots of the dynamics of the vertex model with fixed box boundary condition and anisotropic bond tension. Each cell is colored according to the norm of its elongation tensor and the purple bar in its center indicates the axis of elongation. When the color of the surrounding box becomes red, the mean bond tension $\bar\Lambda_{mn}(t)$ of each bond $\langle mn \rangle$ in the system is gradually activated with the following time-dependence:
\begin{align}
    \bar\Lambda_{mn}(t) = \bar\Lambda^0+ \bar\Lambda^0 \beta_0 \left(1- {\rm e}^{-t/T_{\rm a}} \right)
    \bm{\mathcal{P}}: \hat{\mathbf{H}}_{mn} \, ,
\end{align}
where $\hat{\mathbf{H}}_{mn}$ is the unit nematic tensor of the bond between vertices $m$ and $n$, and the nematic tensor $\bm{\mathcal{P}}$ is along the $y$-axis in this movie.
As a consequence, bond tension is higher for bonds which are more aligned with the vertical ($y$) direction.

\noindent\textbf{Parameters:} $\Lambda_0=0.12$, $\Delta\Lambda=0.06$, $\Gamma=0.04$, $\beta_0=0.5$, $\Sigma^{\rm a}_0=0$, $T_{\rm a}=1$, $\delta t=0.01$.

\subsection{Movie 2: anisotropic bond tension with stress-free boundary conditions}

\noindent\textbf{Description:} Snapshots of the dynamics of the vertex model with stress-free boundary condition and anisotropic bond tension. Each cell is colored according to the norm of its elongation tensor, and the purple bar in its center indicates the axis of elongation. When the color of the surrounding box becomes red, the mean bond tension $\bar\Lambda_{mn}(t)$ of each bond $\langle mn \rangle$ in the system is gradually activated. As a consequence, the higher mean bond tension along the $y$ axis, active T1 transitions are triggered and the box stretches along the $x$ direction.

\noindent\textbf{Parameters:} $\Lambda_0=0.12$, $\Delta\Lambda=0.06$, $\Gamma=0.04$, $\beta_0=0.5$, $\Sigma^{\rm a}_0=0$, $T_{\rm a}=1$, $\delta t=0.01$.

\subsection{Movie 3: anisotropic cell stress with fixed boundary conditions}

\noindent\textbf{Description:} Snapshots of the dynamics of the vertex model with fixed box boundary condition and anisotropic cell stress. Each cell is colored according to the norm of its elongation tensor, and the purple bar in its center indicates the axis of elongation. When the color of the surrounding box becomes red, the anisotropic stress contribution $\Sigma_{ij}(t)$ of the vertex model work function is gradually activated with the following time-dependence:
\begin{align}
    \Sigma_{ij}(t) = \Sigma^{\rm a}_0 \left(1- {\rm e}^{-t/T_{\rm a}} \right) \mathcal{P}_{ij} \, ,
\end{align}
where the nematic tensor $\mathcal{P}_{ij}$ is along the $y$-axis in this movie. 
Note that cells are more contractile along $\mathcal{P}_{ij}$ and therefore elongate in the direction normal to this axis.

\noindent\textbf{Parameters:} $\Lambda_0=0.12$, $\Delta\Lambda=0.06$, $\Gamma=0.04$, $\beta_0=0$, $\Sigma^{\rm a}_0=0.04$, $T_{\rm a}=1$, $\delta t=0.01$.

\subsection{Movie 4: anisotropic cell stress with stress-free boundary conditions}

\noindent\textbf{Description:} \textit{Left panel.} Snapshots of the dynamics of the vertex model with stress-free boundary condition and anisotropic cell stress. Each cell is colored according to the norm of its elongation tensor and the purple bar in its center indicates the axis of elongation. When the color of the surrounding box becomes red, the mean bond tension $\bar\Lambda_{mn}(t)$ of each bond $\langle mn \rangle$ in the system is gradually activated. As a consequence of cell elongation, T1 transitions are triggered and the box stretches along the $y$ direction. \textit{Right panel.} Cumulative shear decomposition of the simulation displayed on the left panel. The color code of the decomposition is the following. Blue represents the total shear, green the contribution from cell elongation to shear, red the contribution of T1 transitions, and purple the contribution due to correlations.

\noindent\textbf{Parameters:} $\Lambda_0=0.12$, $\Delta\Lambda=0.06$, $\Gamma=0.04$, $\beta_0=0$, $\Sigma^{\rm a}_0=0.04$, $T_{\rm a}=1$, $\delta t=0.01$.
